\documentclass[12pt]{article}
\usepackage{amssymb,amsmath,amscd,amsthm}
\usepackage[utf8x]{inputenc}
\usepackage[T2A]{fontenc}
\usepackage{graphicx,psfrag}

\newtheorem{theorem}{Theorem}[section]

\newtheorem{remark}[theorem]{Remark}
\newtheorem{lemma}[theorem]{Lemma}

\newcommand{\m}{\mathcal{M}}
\newcommand{\D}{\mathcal{D}}

\newcommand{\dbar}{\overline{\partial}}

\date{}

\begin{document}
 \author{ Evgeny Lakshtanov\thanks{Department of Mathematics, Aveiro University, Aveiro 3810, Portugal.  This work was supported by Portuguese funds through the CIDMA - Center for Research and Development in Mathematics and Applications and the Portuguese Foundation for Science and Technology (``FCT--Fund\c{c}\~{a}o para a Ci\^{e}ncia e a Tecnologia''), within project UID/MAT/0416/2013 (lakshtanov@ua.pt)}  \and
 Boris Vainberg\thanks{Department
of Mathematics and Statistics, University of North Carolina,
Charlotte, NC 28223, USA. The work was partially supported  by the NSF grant DMS-1714402 and the Simons Foundation grant 527180. (brvainbe@uncc.edu).}}

\title{Global solution of the initial value problem for the focusing Davey-Stewartson II system}

\maketitle

\begin{abstract}
We consider the two dimensional focusing Davey-Stewartson II system and construct the global solution of the Cauchy problem for a dense in $L^2(\mathbb C)$ set of initial data. We do not assume that the initial data is small. So, the solutions may have singularities. We show that the blow-up may occur only on a real analytic variety and the variety is bounded in each strip $t \leq T$.

\end{abstract}


\section{Introduction}
Let $q_0(z), ~ z\!=\!x\!+\!iy, ~ (x,y)\! \!\in \!\mathbb R^2$, be a compactly supported (or fast decaying) sufficiently smooth function. Consider the two dimensional focusing Davey-Stewartson II (DSII) system of equations for unknown functions $q=q(z,t),~\phi=\phi(z,t),  ~(x,y) \in \mathbb R^2, ~ t\geq 0:$
\begin{eqnarray}\nonumber
q_t=2iq_{xy}  - 4q(\overline{\varphi} -\varphi), ~\\ \nonumber
\partial \varphi =\dbar |q|^2, \\
q(z,0)=q_0(z).\label{dsii}
\end{eqnarray}

A smooth solution of (\ref{dsii}) exists for all $t>0$ if $q_0$ is small enough, \cite{bc2, sung1,sung2,sung3,brown}. We will call this solution classical. If $q_0$ is not small, the solution was constructed locally in our previous work \cite{lvds} via the IST (inverse scattering transform) using the $\overline{\partial}$-method that has been generalized in \cite{lnv}, \cite{lbcond} to the case when exceptional points may be present. The solution was obtained in a neighbourhood of any point $(z_0,t_0)$  for generic initial data $q_0$ that depend on the point. The main objectives of this article are to obtain the solution for an arbitrary initial data from a specific set and to get the global solution defined in the whole space, including a description of the set where the solution blows up. It will be shown that the latter set is a real analytic variety that is bounded in every strip $0\leq t\leq T$.

Let us recall that the focusing DSII equation may have  a finite time blow-up (e.g., \cite{ozawa}). While the
uniqueness is known for smooth (in some sense) solutions, see \cite{saut2}, \cite{saut1}, one  has to be careful with the definition of the solution that has singularities. We understand these solutions in the following sense. Let us multiply the initial data $q_0$ by a positive parameter $a \in (0,1]$.  We will show that the classical solution that exists when $a\ll 1$ allows an analytic continuation into a complex neighborhood of $ (0,1]$, and this analytic continuation will be used to single out the solution with singularities when $a=1$.

 Let us mention some recent articles on DSII: \cite{ks},\cite{perry},\cite{perry2}.

\section{The inverse nonlinear Fourier transform} \label{sectmr}

\subsection{The solution of the Cauchy problem, main results}
  Let $q_0(z)\in L^2(\mathbb C)$. Denote
\begin{eqnarray}\label{char1}
Q_0(z)=\left ( \begin{array}{cc}0 &q_0(z) \\
- q_0(z) &  0\end{array} \right ),   \quad  z \in \mathbb C.
\end{eqnarray}
Let $\overline{\partial}=\frac{\partial}{\partial\overline{z}}=\frac{1}{2}\left(\frac{\partial}{\partial x}+i\frac{\partial}{\partial y}\right)$,
and let the $2\times 2$ matrix $\psi(\cdot,k), ~ k \in \mathbb C,$ be a solution of the following problem for the Dirac equation in $\mathbb C:$
\begin{equation}\label{fir}
\frac{\partial \psi}{\partial \overline{z}} = Q_0 \overline{\psi}, \quad \psi(z,k) e^{-i\overline{k}z/2} \rightarrow I, \quad z \rightarrow \infty.
\end{equation}
The corresponding Lippmann-Schwinger equation has the following form:
\begin{eqnarray}\label{19JanA}
\psi(z,k)=e^{i\overline{k}z/2}I+ \int_{\mathbb C} G(z-z',k) Q_0(z') \overline{\psi}(z',k) d\sigma_{z'}, \quad G(z,k)=\frac{1}{\pi}\frac{e^{i\overline{k}z/2 }}{z},
\end{eqnarray}
where $d\sigma_{z'}=d{x'}d{y'}$. Here and below we use the same notation for functional spaces, irrespectively of whether those are the spaces of matrix-valued or scalar-valued functions.  After the substitution,
\begin{equation}\label{27dec2a}
\mu(z,k) = \psi(z,k)e^{-i\overline{k}z/2}, \quad \quad \mu(z,k) -I \rightarrow 0, \quad z \rightarrow \infty,
\end{equation}
equation (\ref{19JanA}) takes the form
\begin{eqnarray}\label{mmmm}
\mu(z,k)=I+ \frac{1}{\pi}\int_{\mathbb C} \frac{e^{i\Re(\overline{k}z) }}{z-z'} Q_0(z') \overline{\mu}(z',k) d\sigma_{z'},
\end{eqnarray}
and becomes Fredholm in $L^q(\mathbb C), ~ q>2 $, after the additional substitution $\nu=\mu-I$ (see, e.g., \cite[lemma 5.3]{music}).

 Solutions $\psi$ of (\ref{19JanA}) are called the {\it scattering solutions}, and the values of $k$ such that the homogeneous equation (\ref{mmmm}) has a non-trivial solution are called {\it exceptional points}. The set of exceptional points  will be denoted by $\mathcal E$. Thus the scattering solution may not exist if $k\in \mathcal E$. Note that the operator in equation (\ref{mmmm}) is not analytic in $k$, and $\mathcal E\subset\mathbb C$ may contain one-dimensional components.
There are no exceptional points in a neighborhood of infinity (e.g., \cite[lemma 2.8]{sung1}, \cite[lemma C]{bu}). Let us choose $A\gg1$ and $k_0 \in \mathbb C$  such that all the exceptional points are contained in the disk
\begin{equation}\label{set28A}
D=\{  k \in \mathbb C: 0 \leq |k| < A\},
\end{equation}
and $k_0 $ belongs to the same disc $ \overline{D}$ and is not exceptional.

The generalized scattering data (an analogue of the scattering amplitude in the standard scattering problem) are defined by the following integral (when the integral converges)
\begin{equation}\label{27dec2}
h_0(\varsigma,k)  =  \frac{1}{(2\pi)^2} \int_{\mathbb C}  e^{-i\overline{\varsigma}z/2}Q_0(z)\overline{\psi}(z,k)d\sigma_z, ~ \varsigma \in \mathbb C,~ k \in \mathbb C\backslash \mathcal E.
\end{equation}
In fact, from the Green formula, it follows that $h_0$ can be determined without using the potential $Q_0$ or the solution $\psi$ of the Dirac equation (\ref{fir}) if the Dirichlet data at  $\partial\Omega$ are known for the solution of (\ref{fir}) in a bounded region $\Omega$ containing the support of $Q_0$:
\begin{equation}\label{2106A}
  h_0(\varsigma,k)  =\frac{-i}{8\pi^2} \int_{\partial\mathcal O}e^{-i\overline{\varsigma}z/2}\overline{\psi}(z,k)dz, ~ \varsigma \in \mathbb C,~ k \not \in \mathcal E.
\end{equation}

Note that $h_0$ is continuous when $k\notin D$ under minimal assumptions on $Q$ \cite{sung1}, \cite{sung2}, and moreover,
\begin{equation}\label{sm}
h_0=h_0(\varsigma,k)\in C^\infty~\quad {\rm when}~~|k|\geq A
\end{equation}
 if $Q$ is bounded and decays faster then any power at infinity. This  follows from the fact that (\ref{mmmm}) admits differentiation in $z$ nd $k$ when $k\notin D$.

{\it The inverse problem} (recovery of $Q$ when $h_0$ is given) was solved using $\dbar-$method in \cite{bc2}, \cite{sung1,sung2,sung3} when the potential is small enough to guarantee the absence of exceptional points. When $\mathcal E\neq \varnothing$, the inverse problem was solved in a generic sense in \cite{lbcond}. The latter results were applied in \cite{lvds} to construct solutions of the focusing DSII system. Let us recall some results obtained in \cite{lvds}.


Consider the space
\begin{equation}\label{170904A}
\mathcal B^s = \left \{ u \in L^s(\mathbb C \backslash D)\bigcap C (D) \right   \}, \quad s>2,
\end{equation}
where $C (D)$ is the space of analytic functions in $D$ with the norm $\|u\|=\sup_D|u|$.
Here and below, we use the same space notation for matrices as for their entries.

Let operator $T_{z}:\mathcal B^s\to \mathcal B^s, ~s>2,$ be defined as follows:
\begin{eqnarray} \nonumber
T_{z} \phi (k)=
\frac{1}{\pi} \int_{\mathbb C\backslash D}  e^{i(\overline{\varsigma}z+\overline{z}{\varsigma})/2} \overline{\phi}(\varsigma) \Pi^o h(\varsigma,\varsigma)  \frac{d\sigma_\varsigma}{\varsigma-k}\\\label{0712GAA}
+\frac{1}{2\pi i} \int_{\partial D} \frac{d\varsigma}{\varsigma - k} \int_{\partial D}
[e^{i/2(\varsigma\overline{z}+\overline{\varsigma'}z)}\overline{\phi^-(\varsigma')} \Pi^o  + e^{i/2(\varsigma-{\varsigma'})\overline{z}}\phi^-(\varsigma')\Pi^d \bold C] \left [{\rm Ln}\frac{\overline{\varsigma'}-\overline{\varsigma} }{\overline{\varsigma'}-\overline{k_0} }h(\varsigma',\varsigma) \overline{d\varsigma'} \right ],
\end{eqnarray}
where $d\sigma_\varsigma=d\varsigma_R d\varsigma_I,~z \in \mathbb C, ~ \phi\in \mathcal B^s$,
$\phi^-$ is the boundary trace of $\phi$ from the interior of $D$,
$\bold C$ is the operator of complex conjugation, $\Pi^oM$ is the off-diagonal part of a matrix $M$, $\Pi^dM$ is the diagonal part.  Let us specify the logarithmic function in (\ref{0712GAA}).
Let us shift the coordinates in $\mathbb C$ and move the origin to the point $\overline{\varsigma'}\in \partial D$. Then we rotate the plane in such a way that the direction of the $x$-axis is defined by the vector from $\overline{\varsigma'}$ to $-\overline{\varsigma'}$.
Then $|\arg(\overline{\varsigma'}-\overline{\varsigma})|< \pi/2, ~\varsigma' \neq \varsigma,$ and $|\arg(\overline{\varsigma'}-\overline{k_0})|\leq \pi/2, $ i.e.,
\begin{equation} \nonumber
\left |\arg \frac{\overline{\varsigma'}-\overline{\varsigma} }{\overline{\varsigma'}-\overline{k_0} } \right | < \pi, \quad \varsigma', \varsigma \in \partial D, ~~\varsigma'\neq \varsigma.
\end{equation}
This defines the values of the logarithmic function uniquely.

It turns out that, after the substitution $w=v-I\in\mathcal B^{s}, ~{s}>2,$ the equation
\[
(I+T_z)v=I
\]
becomes Fredholm in $\mathcal B^{s}$, and the potential $q_0$ can be expressed explicitly in terms of $v$ (see \cite{lnv, lbcond}).

In order to solve the DSII problem  (\ref{dsii}), we apply this reconstruction procedure to a specially chosen scattering data. We start with the scattering data $h_0$ defined by $q_0$ and extend it in time as follows:
\begin{equation}\label{2506B}
h(\varsigma,k,t):=e^{-t(k^2-\overline{\varsigma}^2)/2}\Pi^o h_0(\varsigma,k) +e^{-t(\overline{k}^2-\overline{\varsigma}^2)/2}\Pi^d h_0(\varsigma,k),~\varsigma\in\mathbb C, ~ k \in \mathbb C\backslash \mathcal E, ~ t \geq 0.
\end{equation}
For $t \geq 0$, we define the operator
\begin{eqnarray} \nonumber
T_{z,t} \phi (k)=
\frac{1}{\pi} \int_{\mathbb C\backslash D}  e^{i(\overline{\varsigma}z+\overline{z}{\varsigma})/2} \overline{\phi}(\varsigma) \Pi^o h(\varsigma,\varsigma,t)  \frac{d\sigma_\varsigma}{\varsigma-k}\\\label{0712G}
+\frac{1}{2\pi i} \int_{\partial D} \frac{d\varsigma}{\varsigma - k} \int_{\partial D}
[e^{i/2(\varsigma\overline{z}+\overline{\varsigma'}z)}\overline{\phi^-(\varsigma')} \Pi^o  + e^{i/2(\varsigma-{\varsigma'})\overline{z}}\phi^-(\varsigma')\Pi^d \bold C] \left [{\rm Ln}\frac{\overline{\varsigma'}-\overline{\varsigma} }{\overline{\varsigma'}-\overline{k_0} }h(\varsigma',\varsigma,t) \overline{d\varsigma'} \right ].
\end{eqnarray}

\begin{theorem}\label{mthm1}(\cite{lvds}) Let $q_0(\cdot)$ a function with  compact support. Assume that it is 6 times differentiable in $x$ and $y$. Alternatively,
this condition can be replaced\footnote{This fact was not mentioned in the paper, but it can be easily checked} by the superexponential decay of $q_0$:
\begin{equation}\label{adad}
\lim_{z\to \infty}e^{\widetilde{A}|z|}\partial_x^i \partial_y^j q_0(z)=0~~ for ~each ~~\widetilde{A}>0,~ i+j\leq 6.
\end{equation}
  Then, for each  $s>2$, the following statements are valid.
\begin{itemize}
\item
The operator $T_{z,t}$ is compact in $\mathcal B^{ s}$ for all $z \in \mathbb C, ~t \geq 0,$ and depends continuously on $z$ and $t \geq 0$.
The same property holds for its first derivative in time and all the derivatives in $x,y$ up to the third order, where the derivatives are defined in the norm convergence. The function $T_{z,t} I$ belongs to $\mathcal B^{ s}$ for all $t \geq 0$.

  \item Let the kernel of $I+T_{z,t}$ in the space $\mathcal B^{s}$ be trivial for $(z,t)$ in an open {or  half open\footnote{We will say that a set $\omega$ of points $(z,t)$ in $ \mathbb R^3_+=\mathbb R^3\bigcap\{t\geq 0\}$ is half-open if $ \omega $ contains points where $t=0$ and, for each point $(z_0,0)\in\omega$, there is a ball $B_0$ centered at this point such that $B_0\bigcap\{t\geq 0\}\subset\omega$.
}} set $\omega\subset  \mathbb R^3$. Let $v_{z,t}=w_{z,t}+I$, where $w_{z,t}\in \mathcal B^{s}$ is the solution of the equation
      \begin{equation}\label{inteq}
 (I+T_{z,t})w_{z,t} =-T_{z,t} I.
 \end{equation}
Then functions $q,\varphi$ defined by
\begin{eqnarray}\nonumber
\left (  \!\!\! \begin{array}{cc} \varphi(z,t) & q(z,t) \\ - q(z,t) & \varphi(z,t) \end{array} \!\! \! \right )
:=\frac{-i}{2}(\Pi^o + \dbar \Pi^d) \left (
 \frac{1}{\pi} \int_{\mathbb C\backslash D}  e^{i(\overline{\varsigma}z+\overline{z}{\varsigma})/2} \overline{v_{z,t}}(\varsigma) \Pi^oh(\varsigma,\varsigma,t)  {d\sigma_\varsigma}  \right .\\ \left .
-\frac{ 1}{2\pi i} \int_{\partial D}\! {d\varsigma} \!\int_{\partial D}
[e^{i/2(\varsigma\overline{z}+\overline{\varsigma'}z)}\overline{v_{z,t}^-(\varsigma')} \Pi^o  - e^{i/2(\varsigma-{\varsigma'})\overline{z}}v_{z,t}^-(\varsigma')\Pi^d \bold C] \left [{\rm Ln}\frac{\overline{\varsigma'}-\overline{\varsigma} }{\overline{\varsigma'}-\overline{k_0} }h(\varsigma',\varsigma,t) \overline{d\varsigma'} \right ] \!\right ),
\label{1112A}
\end{eqnarray}
 satisfy all the relations (\ref{dsii}) in the classical sense when $(z,t)\in\omega$,. In particular, $q(z,0)=q_0(z)$.
\item
 Consider a set of initial data $aq_0(z)$ that depend on  $a \in (0,1]$. Then equation (\ref{inteq}) with $Q^0$ replaced by $a Q^0$ ($Q^0$ is fixed) is uniquely solvable for almost every $(z, t, a)\in\mathbb R^2 \times \mathbb R^+ \times (0,1]$. Moreover,\footnote{that statement can be found in \cite[lemma 5.1]{lbcond}} for each $(z,t)$, the solution of (\ref{inteq}) is meromorphic in $a\in[0,1]$ and has at most a finite set of poles $a=a_j(z,t)$.
   \end{itemize}
\end{theorem}
{\bf Remark.} All the exceptional points are located in a disk whose radius depends only on the norm of $aq_0$. Hence $D$ and $k_0$ can be chosen independently of $a\in [0,1]$ (see \cite[Lemma 5.1]{lbcond}). From now on, we assume that the disk $D$ is fixed and contains the exceptional points for all the potentials $aq_0,~a\in[0,1]$.

 In order to state the main results of the present paper, we need to recall the construction (e.g., \cite{sung1}) of the global solution of (\ref{dsii}) when $q_0$ is small. The latter expression ("$q_0$ is small") will be used below only for problem (\ref{dsii}) with initial data $aq_0$ where $q_0$ is infinitely smooth and satisfies (\ref{adad}), and $0<a\ll 1$. Let us recall that the scattering problem (\ref{fir}) and the Lippmann-Schwinger equation (\ref{19JanA}) are uniquely solvable for all $k$ when $q_0$ is small, i.e., there are no exceptional points in this case and $h_0(k, k )$  is defined for all the values of $k$. Operator $T_{z,t}$ is needed only with $D=\emptyset$ if $q_0$ is small. Hence, only the first term is present in the right-hand side of  (\ref{0712G}). Moreover, $\|T_{z,t}\|<1$ when $q_0$ is small, and therefore equation (\ref{inteq}) is uniquely solvable for all $z\in \mathbb C, ~t\geq 0$.
Then $(q,\phi)$  given by (\ref{1112A}) with $D=\emptyset$ is a smooth global solution of problem (\ref{dsii}) with the small initial data. We will call this solution {\it classical}. It exists under a weaker assumption on the decay of $q_0$ than in Theorem \ref{mthm1}.

We will consider analytic continuations of functions $h_0, q_0$, and we need some notation. Let $\gamma=(\gamma_1,\gamma_2)\in \mathbb C^2$ and let $A_\gamma:\mathbb C\to\mathbb C^2$
be the map defined by $A_\gamma z=A_\gamma (x+iy)=(x+\gamma_1,y+\gamma_2)\in\mathbb C^2$, i.e., the map $A_\gamma$ shifts real points $x,y$ into complex planes.
If a function $f=f(z)$ is analytic in $(x,y)$, then $B_\gamma f(z):=f(A_\gamma z)$ is the value of the analytic continuation of $f$ at point $A_\gamma z$.
 We will use notation $A_\sigma', B_\sigma',~\sigma\in\mathbb C^2,$ for the same operations applied to a function of $k\in\mathbb C$, and $A_\eta'', B_\eta'',~\eta\in\mathbb C^2,$ if they are applied to a function  of $\varsigma\in\mathbb C$.


The main result of the paper is obtained under the following condition on the initial data that must hold for large enough $R$:

{\bf Condition $Q(R)$.  } The initial data $q_0$ admits analytic continuation in $(x,y )$ and, for a given $R>0$, there exist a $C=C(R)$ such that
 \[
 |B_\gamma q_0(z)|\leq Ce^{-R|z|}, \quad z\in \mathbb C, ~~{\rm when} ~~ |\gamma| \leq R.
 \]
\begin{remark}\nonumber  Clearly, linear combinations of Gaussian functions satisfy Condition $Q(R)$ for all $R>0$, and it was shown in \cite{CB, cr1} that these combinations form a dense set in $L^2(\mathbb C)$.
\end{remark}
We will show that Condition $Q(R)$ implies a similar behavior of the scattering data, i.e., the validity of the following assumption.

{\bf Condition $H(R)$.} For a given $R>0$, the estimate
\[
|h_0(\varsigma,\varsigma)|\leq e^{-R|\varsigma|} \quad {\rm as}\quad |\varsigma|\to \infty
\]
holds, and there exists  $C=C(R)$  and $a_0(R)$ such that the scattering data $h_0(\varsigma, \varsigma)$ for the potential $aq_0$ with $a\in(0,a_0)$ admits analytic continuation $B_\eta''h_0(\varsigma,\varsigma),~|\eta| \leq R,$ with respect to variables $\varsigma_i$, and
\begin{equation} \nonumber
|B_\eta''h_0(\varsigma,\varsigma)|\leq C(R)e^{-R|\varsigma|}, \quad \varsigma\in \mathbb C, ~~{\rm when} ~~ |\eta| \leq R.
\end{equation}

We will show that there is a duality between these two conditions. To be more exact, the validity of $Q(R)$ implies the validity of $H(R-\varepsilon)$. Conversely, if the initial data is small, Condition $H(R)$ holds, and $h_0$ is extended in time according to (\ref{2506B}), then the potential $q(z,t)$ that corresponds to the extended data $h(\varsigma,\varsigma,t)$ satisfies Condition $Q(R)$ with a smaller $R$ that depends on $t$. These results will be obtained in the next section. Note that they are an analogue of similar results of L. Sung (\cite[Cor. 4.16  ]{sung2}) who established a duality of the non-linear Fourier transform in the Schwartz class. We need a refined result to study the more complicated form (\ref{0712G}) of operator $T_{z,t}$  that appears in the presence of exceptional points.

 Below is the main statement of the present paper.

\begin{theorem}\label{t111}
Let us fix an arbitrary disk $D$ containing all the exceptional points for the potentials $aq_0(z), ~a\in[0,1]$. Let Condition $Q(R)$ hold with $R>(1+2T)A$, where $A$ is the radius of the disk $D$. Then

1) for each point $(z,t),~0\leq t\leq T$, the classical solution  $(q,\phi)$ of problem (\ref{dsii}) with the initial data $~aq_0$ and small enough $a>0$ admits a meromorphic continuation with respect to $a$ in a neighborhood of the segment $[0,1]$. This meromorphic continuation is given by (\ref{1112A}) with an arbitrary choice of the disk $ D$ and an arbitrary choice of point $k_0\in \partial D$\footnote{all the exceptional points are inside $D$, i.e., all the points on $\partial D$ are non-exceptional. }.

2) when $a=1$, the analytic continuation of $(q,\phi)$ is infinitely smooth and satisfies (\ref{dsii}) everywhere, except possibly a set $S$ that is bounded in the strip $0\leq t\leq T,~(x,y)\in R^2,$ and is such that $S_t=S\bigcap\{t=const\}$ is a bounded 1D real analytic variety.

\end{theorem}

{\bf Remark.} The theorem implies that the local solutions found in Theorem \ref{mthm1} are analytic continuations in $a$ of the global classical solutions (under the assumption that condition $Q(R)$ holds). At the same time, the theorem does not prohibit the solution from blowing up at an arbitrarily small time $t>0$ (see the recent
paper \cite{korp} and citations there on instantaneous blow-ups). We can't say anything about relation between our global solution and local solutions found in \cite{saut2}.

Two important technical improvements of the previous results will be used in the proof of Theorem \ref{t111}. First, we will show that the Hilbert space $\mathcal B^2$ can be used in Theorem~\ref{mthm1} instead of the Banach space $\mathcal B^s,~s>2$. The space $\mathcal B^2$ is defined as follows:
\begin{equation}\label{170904AZ}
\mathcal B^2 = \left \{ u \in \left (L^2(\mathbb C \backslash D) \oplus \mathbb C^1 \right) \bigcap L^2_+ (\partial D) \right   \}.
\end{equation}

Here $\mathbb C^1$ is the one-dimensional space of functions of the form $\frac{c\beta(k)}{k}$,
where $c$ is a complex constant, $\beta \in C^\infty$ is a fixed function that vanishes in a neighbourhood of the disk $D$ and equals one in a neighbourhood of infinity. By $L^2_+ (\partial D)$ we denote the space of analytic functions $u=\sum_{n\geq 0}c_nz^n$ in $D$ with the boundary values in $L^2(\partial D)$ and the norm
$$
\|u\|_{L^2_+ (\partial D)} = \left ( \sum_{n \geq 0} A^{2n}|c_n|^2 \right )^{1/2},
$$
where $A$ is the radius of the disk $D$.

Secondly, we will simplify the form of the operator $T_{z,t}$ by writing the second term in (\ref{0712GAA}) and (\ref{0712G}) without the logarithmic factor. We also will allow $k_0$ to be on $\partial D$, and not necessarily in $D$, and show that formula  (\ref{0712G}) in the latter case can be written as
\begin{eqnarray} \nonumber
T_{z,t} \phi (k)=
\frac{1}{\pi} \int_{\mathbb C\backslash D}  e^{i(\overline{\varsigma}z+\overline{z}{\varsigma})/2} \overline{\phi}(\varsigma) \Pi^o h(\varsigma,\varsigma,t)  \frac{d\sigma_\varsigma}{\varsigma-k}-\\\label{0712GNNN}
i \int_{\partial D} \frac{d\varsigma}{\varsigma - k} \int_{\widehat{k_0,\varsigma}}
[e^{i(\varsigma\overline{z}+\overline{\varsigma'}z)/2}\overline{\phi^-(\varsigma')} \Pi^o  + e^{i(\varsigma-{\varsigma'})\overline{z}/2}\phi^-(\varsigma')\Pi^d \bold C] \left [h(\varsigma',\varsigma,t) \overline{d\varsigma'} \right ],
\end{eqnarray}
where $\widehat{k_0,\varsigma}$ is the arc on $\partial D$  between points $k_0$ and $\varsigma$ with the counter clock-wise direction on it.

 The following two difficulties were resolved in the paper. We show that if one starts with a small potential $q_0$ and its scattering data $h_0(\varsigma,\varsigma)$, and  extends $h_0(\varsigma,\varsigma)$ in time according to (\ref{2506B}), then the solution $q(z,t)$ of the inverse scattering problem with the scattering data (\ref{2506B})  decays exponentially at infinity, and the scattering data (\ref{27dec2}) for this potential $q(z,t)$ coincides with the scattering data  $h(\varsigma,\varsigma,t)$ from which the potential was obtained (this will be done in the next section). Another difficulty concerns the proof of the invertibility of operator $I+T_{z,t}$ for large $|z|$ in spite of the exponential growth of the integrands in the second terms of (\ref{0712G}) and (\ref{0712GNNN}) as $|z|\to\infty$ (see section 5).

\section{Exponential decay of the scattering data and of $q(z,t)$}

 \begin{lemma}\label{l1a}
 Let
\begin{equation} \nonumber
 I(z)=\int_\mathbb C\frac{f(z_1)}{z-z_1}d\sigma_{z_1},\quad  J(z)=\int_\mathbb C\frac{f(z_1)}{\overline{z}-\overline{z}_1}d\sigma_{z_1}, \quad z\in\mathbb C,
 \end{equation}
 where $f(z)$ is analytic in $(x,y)$, and
 \[
 |f(A_\gamma z)|,|\nabla_\gamma f(A_\gamma z)|\leq \frac{C(\gamma)}{1+x^2+y^2}.
 \]
  Then $I(z), J(z)$ admit analytic continuation in $(x,y)$, and
 \[
B_\gamma I( z)=\int_\mathbb C\frac{f(A_\gamma z_1)}{z-z_1}d\sigma_{z_1}, \quad B_\gamma J( z)=\int_\mathbb C\frac{f(A_\gamma z_1)}{\overline{z}-\overline{z}_1}d\sigma_{z_1}.
 \]
 \end{lemma}
{\bf Proof.} Let us rewrite $I(z)$ in the form
\[
I(z)=-\int_\mathbb C\frac{f(z+z_1)}{z_1}d\sigma_{z_1}.
\]
This immediately implies that $I(z)$ is analytic in $(x,y)$, and
\[
B_\gamma I(z)=\int_\mathbb C\frac{B_\gamma f(z+z_1)}{-z_1}d\sigma_{z_1}=\int_\mathbb C\frac{ f(x+\gamma_1+x_1,y+\gamma_2+y_2)}{-z_1}d\sigma_{z_1}=\int_\mathbb C\frac{f(A_\gamma z_1)}{z-z_1}d\sigma_{z_1}.
\]
The statement for $J$ can be proved absolutely similarly.

\qed

Let us provide some examples of analytic continuations of functions from $\mathbb C$ into $\mathbb C^2$: (1) If $f(z)=z=x+iy$, then $B_\gamma f(z)=x+\gamma_1+i(y+\gamma_2)=z+\gamma', ~ \gamma'=\gamma_1+i\gamma_2\in \mathbb C$. (2) If $f(z)=\overline{z}=x-iy$, then $B_\gamma f(z)=x+\gamma_1-i(y+\gamma_2)=\overline{z}+\gamma'', ~ \gamma''=\gamma_1-i\gamma_2\in \mathbb C$ (note that $\gamma'\neq \overline{\gamma''}$ since $\gamma_i$ are complex.) (3) if $f(z)=\Re(k\overline{z})=k_1x+k_2y$, then $B_\gamma f(z)=\Re(k\overline{z})+k_1\gamma_1+k_2\gamma_2$ and
 $B_\sigma \Re(k\overline{z})=\Re(k\overline{z})+\sigma_1x+\sigma_2y$.

\begin{lemma}\label{xzx}
 Let the potential be $aq_0(z)$ where $q_0$ satisfies Condition $Q(R)$ for some $R>0$. Then there exists $a_0=a_0(R)$ such that function $\overline{\mu}=\overline{\mu}(z,k)$ defined by (\ref{19JanA}) via the solution of the Lippmann-Schwinger equation with the potential $aq_0,~a\in(0,a_0),$  admits analytic continuation to $\mathbb C^4$ with respect to variables $x,y, k_1,k_2$, and
 \begin{equation}\label{mumu}
 |B_\sigma' B_\gamma\overline{\mu}(z,k)|<C(R,\varepsilon)  \quad { when}~~ z,k\in\mathbb C,~~~|\sigma|,|\gamma|\leq R-\varepsilon, ~~a<a_0.
\end{equation}
 The statement remains valid if $a=1$, but $|k|\geq \rho(R)$ with large enough $\rho$.
 \end{lemma}
{\bf Proof.} We will prove the statement of the lemma for the component $\mu_{11}$ of the matrix $\mu$. Other components can be treated similarly.
Let us iterate equation (\ref{mmmm}). The following equation is valid for the first component:
\begin{equation}\label{2006C}
\overline{\mu_{11}}=1+
\frac{1}{\pi^2} \int_{\mathbb C} d\sigma_{z_1} \int_{\mathbb C} d\sigma_{z_2} \frac{e^{i \Re(k\overline{z}_1)}}{\overline{z}-\overline{z_1}} \overline{Q}_{12}(z_1) \frac{e^{-i \Re(k\overline{z}_2)}}{z_1-z_2} {Q_{21}}(z_2) \overline{\mu_{11}}(z_2,k),
\end{equation}
where $Q_{21}$ and $Q_{12}$ are entries of the matrix $Q_0$. Denote $Q=Q_{12}=-Q_{21}$. Assume that the analytic continuation $ B_\gamma\overline{\mu}_{11}$ exists. Then from Lemma \ref{l1a} and (\ref{2006C}) it follows that
 \[
 B_\gamma\overline{\mu}_{11}=1-\frac{1}{\pi^2} \int_{\mathbb C} d\sigma_{z_1}  \frac{B_\gamma e^{i \Re(k\overline{z}_1)}}{\overline{z}-\overline{z_1}}B_\gamma \overline{Q}(z_1)B_\gamma \int_{\mathbb C} d\sigma_{z_2}\frac{e^{-i \Re(k\overline{z}_2)}}{z_1-z_2} {Q}(z_2) \overline{\mu_{11}}(z_2,k)
 \]
 \[
 =1-\frac{1}{\pi^2} \int_{\mathbb C} d\sigma_{z_1}  \frac{ e^{i \Re(k\overline{z}_1)+i<k,\gamma>}}{\overline{z}-\overline{z_1}}B_\gamma \overline{Q}(z_1)\int_{\mathbb C}\frac{e^{-i \Re(k\overline{z}_2)-i<k,\gamma>}}{z_1-z_2} B_\gamma {Q}(z_2)B_\gamma  \overline{\mu_{11}}(z_2,k)d\sigma_{z_2}
 \]
 \[
 =1-\frac{1}{\pi^2} \int_{\mathbb C} d\sigma_{z_1}  \frac{ e^{i \Re(k\overline{z}_1)}}{\overline{z}-\overline{z_1}}B_\gamma \overline{Q}(z_1)\int_{\mathbb C}\frac{e^{-i \Re(k\overline{z}_2)}}{z_1-z_2} B_\gamma {Q}(z_2)B_\gamma  \overline{\mu_{11}}(z_2,k)d\sigma_{z_2}.
 \]
 Hence, if the analytic continuation $\Psi:=B_\sigma' B_\gamma\overline{\mu}_{11}$ exists, then it satisfies the equation
\begin{equation*}
\Psi(z,k)=1-
\frac{1}{\pi^2} \int_{\mathbb C} d\sigma_{z_1} \frac{e^{i \Re(k\overline{z}_1)+i<\sigma,z_1>}}{\overline{z}-\overline{z_1}} B_\gamma\overline{Q}(z_1) \int_{\mathbb C}\frac{e^{-i \Re(k\overline{z}_2)-i<\sigma,z_2>}}{z_1-z_2} B_\gamma{Q}(z_2)\Psi(z_2,k)d\sigma_{z_2}.
\end{equation*}
Denote by $K^\pm=K^\pm_{k,\sigma,\gamma}$ the integral operators given by the exterior and interior integrals above, respectively. Their norms in the space  $L^\infty(\mathbb C)$  can be estimated from above by the norms of the potential (see \cite{sung1}):
\[
\|K^- \|< C(\|e^{-i<\sigma,z_2>}B_\gamma Q(z_2)\|_{L^p(\mathbb C)}+\|e^{-i<\sigma,z_2>}B_\gamma Q(z_2)\|_{L^q(\mathbb C)}), \quad 1<p<2<q<\infty,
\]
and a similar estimate is valid for $K^+$. Thus the assumption $a_0\ll 1$ and Condition Q imply that $\|K^\pm\|<1$, $\Psi$ exists, and
\[
 |\Psi|<C(R)  \quad {\rm when}~~ |\Im\sigma|,|\gamma|\leq R, ~a<a_0.
 \]
 Moreover, the derivatives of $K^\pm$ with respect to complex variables $\sigma_i,\gamma_j$ also have small norms, i.e., $\Psi =\Psi(z,k,\sigma,\gamma)$ is analytic in $(\sigma_1,\sigma_2,\gamma_1, \gamma_2)$.  One can easily see that $\Psi =\Psi(z+\gamma,k+\sigma)$. Hence $\Psi$ is the analytic continuation of $\overline{\mu}$. The proof of (\ref{mumu}) is complete.

 In order to prove the statement of Lemma \ref{xzx} concerning $a=1$, one needs only to show that operator $K:=K^+K^-$ and its derivatives in $\sigma_i,\gamma_j$ are small (less than one) as $|k|\to\infty$. This can be done by a standard procedure: one splits $K$ into two terms $K=K_1+K_2$, where $K_1$ is obtained by adding the factor $\alpha(\frac{z_1-z}{\varepsilon})\alpha(\frac{z_1-z_2}{\varepsilon})$ in the integral kernel of $K$. Here $\alpha=\alpha(z)$ is a cut-off function that is equal to one when $|z|<1$ and vanishes when $|z|>2$. Then $\|K_1\|\to 0$ as $\varepsilon\to 0$, and $\|K_2\|=O(|k|^{-1})$ as $|k|\to\infty$. The latter can be shown by appropriate integration by parts in $x_1,y_1$.

\qed

\begin{theorem}\label{t222}
 If  Condition $Q(R)$ holds for some $R>0$, then Condition $H(R-\varepsilon)$ holds for each $\varepsilon > 0$.
\end{theorem}

{\bf Proof.} Recall that
\begin{equation} \nonumber
h_0(\varsigma,\varsigma)  =  \frac{1}{(2\pi)^2} \int_{\mathbb C}  e^{-i\Re(\overline{\varsigma}z)}Q_0(z)\overline{\mu}(z,\varsigma)d\sigma_z, ~ k,\varsigma \in \mathbb C.
\end{equation}
 We shift the complex plane $\mathbb C$ in the integral above by vector $\gamma=-i\frac{(\varsigma_1,\varsigma_2)}{|\varsigma|}(R-\varepsilon)$, and then apply operator $B_\eta$. This leads to
\[
|B_\eta''h_0|< \frac{1}{(2\pi)^2} \int_{\mathbb C} \left | e^{-i(<\eta,z>+<\varsigma,\gamma>+<\eta,\gamma>)}Q_0(A_\gamma z)B_\gamma B_\eta''\overline{\mu}(z,\varsigma)\right | d\sigma_z.
\]
It remains to use Lemma \ref{xzx} and Condition $Q(R)$.

\qed

\begin{theorem}\label{tttt}
Let Condition $Q(R)$ hold for some $R>0$ and let the scattering data $h_0$ be defined by the potential $aq_0,~0<a<a_0(R),$ where $a_0(R)$ is defined in Lemma \ref{xzx}. Then the time dependent scattering data $h(\varsigma,\varsigma,t),~0\leq t\leq T,$ given by (\ref{2506B}), admits an analytic continuation in $(\varsigma_1,\varsigma_2)$, and
\[
|B_\eta''h(\varsigma,\varsigma,t)|\leq C(R,\varepsilon)e^{(-\frac{R}{1+2T}+\varepsilon)|\varsigma|}, \quad |\eta|\leq \frac{R}{1+2T}.
\]
The statement remains valid if $a=1$, but $|\varsigma|>\rho,$ where $\rho=\rho(R)$ is large enough.
\end{theorem}
{\bf Proof.} The statement follows immediately from Theorem \ref{t222} and formula (\ref{2506B}). One needs only to combine the upper bound $Ce^{(-R+\varepsilon)|\varsigma|}$ for the analytic continuation of $h_0$ obtained in Theorem \ref{t222} with the upper bound $Ce^{\frac{2TR}{1+2T}|\varsigma|}$ for the time-dependent factor in (\ref{2506B}).

\qed

Let us recall again the procedure to obtain the classical solution of the focusing DSII equation with initial data $aq_0$ and a very small $a$ such that there are no exceptional points. As the first step, one needs to solve the equation $(I+T_{z,t})v=I$, where $T_{z,t}$ is given by (\ref{0712G}) with $D=\emptyset$, i.e., the equation for $v=v_{z,t}$ has the form
\begin{equation}\label{inv1}
v_{z,t}(k)+\frac{1}{\pi} \int_{\mathbb C}  e^{i(\overline{\varsigma}z+\overline{z}{\varsigma})/2} \overline{v_{z,t}}(\varsigma) \Pi^o h(\varsigma,\varsigma,t)  \frac{d\sigma_\varsigma}{\varsigma-k}=I, \quad w_{z,t}(\cdot)=v_{z,t}(\cdot)-I\in\mathcal B^{s}.
\end{equation}
Then the solution of the focusing DSII equation with initial data $aq_0$ is given by (\ref{1112A}). In particular,
\begin{equation}\label{inv2}
q(z,t)=\frac{1}{2\pi i} \int_{\mathbb C}  e^{i(\overline{\varsigma}z+\overline{z}{\varsigma})/2} (\overline{v_{z,t}})_{11}(\varsigma) h_{12}(\varsigma,\varsigma,t)  {d\sigma_\varsigma}.
\end{equation}
\begin{theorem}\label{ttttl}
Let Condition $Q(R)$ hold for $q_0$, and let the potential $q(z,t),~0\leq t\leq T,$ in (\ref{inv2}) be constructed from the initial data $aq_0(z)$ with $0<a<a_1\ll 1$. Then there exists $a_1=a_1(R,T)$ such that Condition $Q(\frac{R}{1+2T}-\varepsilon)$ holds  for the potential (\ref{inv2}) for all  $~t\in [0,T]$.
\end{theorem}

{\bf Proof.} There is a complete duality (e.g. \cite[Th. 4.15]{sung2}) between the nonlinear Fourier transform given by (\ref{mmmm}), (\ref{27dec2}) and the inverse transform (\ref{inv1}), (\ref{inv2}). Function $h$ in (\ref{inv1}) plays the role of the potential $Q_0$ in (\ref{mmmm}). Theorem \ref{tttt} implies that the Condition
$Q(R')$ holds for $h$ with $R'=\frac{R}{1+2T}-\frac{\varepsilon}{2}$. From Lemma \ref{xzx} applied to (\ref{inv1}) instead of (\ref{mmmm}), it follows that $v$ has the same properties as the properties of $\mu$ established in Lemma \ref{xzx}. One needs only to take $a$  small enough to guarantee that the analogues of operators $K^\pm$ have norms that do not exceed one. Then
\begin{equation} \nonumber
 |B_\sigma' B_\gamma\overline{v}(z,k)|<C(R',\varepsilon)  \quad {\rm when}~~ z,k\in\mathbb C,~~~|\sigma|,|\gamma|\leq R'-\frac{\varepsilon}{2}, ~~a\ll 1.
\end{equation}
Then the statement of the theorem can be obtained similarly to the proof of Theorem~\ref{t222}, i.e., by using the shift of the complex plane $\mathcal C$ in (\ref{inv2}) by the vector $\eta=i\frac{(x,y)}{|z|}(R'-\frac{\varepsilon}{2})$.

\qed

\section{Proof of the first statement of Theorem \ref{t111}}

 Consider problem (\ref{dsii}) with $q_0$ replaced by $aq_0,~a \in (0,1]$. Let $ D$ be a disk containing all the exceptional points for problems (\ref{fir}), (\ref{19JanA}) for all $a\in (0,1]$.  Let  $k_0 \in \partial D$  be a non-exceptional point for all $a\in (0,1]$.  We will use notation $v^{1}$ for the solution of  (\ref{inteq}) and $(q^{1},\varphi^{1})$  for the pair  defined by (\ref{1112A}) when the operator $T_{z,t}$ is defined using the disk $D$. We preserve the notations $v, ~(q,\varphi)$ for the same objects when there are no exceptional points and $D=\emptyset$. Since $q^{1},\varphi^{1}$ are meromorphic in $a$ in a neighbourhood of $(0,1]$ (see Theorem \ref{mthm1}), the first statement of Theorem \ref{t111} will be proved if we show that $(q^{1},\varphi^{1})= (q,\varphi)$ when $a>0$ is small and $t>0$.


 From (\ref{27dec2a}),  (\ref{27dec2}) and Condition $Q(R)$ with $R>(1+2T)A>A$, it follows that the scattering data $h_0=h_0(\varsigma, k )$ is defined for all the potentials $aq_0$ when $|\varsigma|, |k|\leq A$ (i.e., $\varsigma, k\in \overline{D}$) and also for all $\varsigma=k$.
We define $h(\varsigma,k,t)$ (extension of $h_0$ in $t$) according to (\ref{2506B}). Let  $v=v_{z,t}=w_{z,t}+I$, where $w_{z,t}\in \mathcal B^{s}, s>2$, is the solution of (\ref{inteq})  with $T_{z,t}$ given by (\ref{0712G}) with $D=\emptyset$ (i.e., the right-hand side in (\ref{0712G}) contains only the first term, see equation (\ref{inv1})). Then  $(q,\phi)$ given by (\ref{1112A}) with $D=\emptyset$ solves the DSII equation (\ref{dsii}) (see \cite{FS}), and
 \begin{equation}\label{psiv}
 \psi=\psi(z,k,t):=\Pi^d \overline{v}e^{i\overline{k}z/2} + e^{-i\overline z k/2} \Pi^o v, ~ \varsigma,k \in \mathbb C, ~ t \geq 0,
\end{equation}
is the solution of the scattering problem (\ref{fir}) (and the Lippmann-Schwinger equation (\ref{19JanA})) with the potential $Q_t(z)=\left ( \begin{array}{cc}0 &q(z,t) \\
- q(z,t) &  0\end{array} \right )$ instead of $Q_0$.

Consider now the scattering data
 \begin{equation}\label{170204A}
 \widehat{h}(\varsigma,k,t):=\frac{1}{(2\pi)^2} \int_{\mathbb C}  e^{-i\overline{\varsigma}z/2}Q_t(z)\overline{\psi}(z,k,t)d\sigma_z
 \end{equation}
defined by the solution $\psi$ of the Lippmann-Schwinger equation (\ref{19JanA}) with the potential $Q_t(z)$. If $0\leq t\leq T$, then from Theorem \ref{ttttl} (it is assumed there that $R>(1+2T)A$) it follows that integral (\ref{170204A}) converges when $|\varsigma|, |k|\leq A$ (i.e., $\varsigma, k\in \overline{D}$) and when $\varsigma=k$. Moreover, $\widehat{h}(\varsigma,k,t)=\widehat{h}(k+\alpha,k,t)$ is an anti-analytic continuation of $ \widehat{h}(k,k,t) $ in $\alpha$. We will prove that $\widehat{h}$ coincides with the scattering data $h(\varsigma,k,t)$ defined in (\ref{2506B}).
We also will prove that there exists an analytic in $k$ function $\widehat{v}_1^+=\widehat{v}_1^+(k,t), ~ k\in D$, such that
\begin{equation}\label{170204B}
\left (v-\widehat{v}^+_1)\right|_{\varsigma \in\partial D}=\int_{\partial D}
[e^{i/2(\varsigma\overline{z}+\overline{\varsigma'}z)}\overline{\widehat{v}_1^+(\varsigma')} \Pi^o  - e^{i/2(\varsigma-{\varsigma'})\overline{z}}\widehat{v}_1^+(\varsigma')\Pi^d \bold C] \left [{\rm Ln}\frac{\overline{\varsigma'}-\overline{\varsigma} }{\overline{\varsigma'}-\overline{k_0^1} }\widehat{h}_t(\varsigma',\varsigma) d\varsigma' \right ].
\end{equation}
From these two facts and the $\overline{\partial}$-equation (see \cite{sung1})
\begin{equation}\label{dbar}
\frac{\partial}{\partial \overline{k}} v(z,k,t)= e^{i(\overline{k}z+\overline{z}k)/2} \overline{v}(z,k,t) \Pi^o h(k,k,t), \quad k\in \mathbb C \backslash D,
\end{equation}
 it follows   (see \cite[Lemma 3.3]{lbcond}) that the function
\begin{equation}\label{pspr}
v'(z,k) := \left \{ \begin{array}{l}
v(z,k), \quad k\in \mathbb C \backslash D, \\
\widehat{v}_1^+(z,k), \quad  k\in D,
\end{array}
\right .
\end{equation}
satisfies the integral equation (\ref{inteq}), where operator $T_{z,t}$ is constructed using the scattering data $\widehat{h}$. Equation (\ref{inteq}) has a unique solution when $a$ is small enough. Under the assumption that $\widehat{h}=h$, we have $v^1 \equiv v'$. Therefore $v^1(z,k)=v(z,k)$ when $ k \in \mathbb C \backslash D$. Solution $(q,\phi)$ of the DSII equation can be determined via the asymptotics of $v$ at large values of $k$  (e.g., \cite[(1.17)]{sung1}, \cite[Lemma 3.3]{lvds}). Hence $(q^1,\phi^1)= (q,\phi)$  for small $a$. Thus the first statement of the theorem will be proved as soon as we show that  $\widehat{h}=h, ~t>0$, and that $\widehat{v}^+$ exists.

{\it Justification of the equality $\widehat{h}=h, ~t>0$.} Everywhere below, till the end of the section, we omit mentioning the parameter $a$ and assume that the initial data $q_0$ is small. Let us recall (see \cite [Lemmas 4.1, 4.2]{lvds}) that the symmetry of the matrix $Q_0$ (see (\ref{char1})) implies that $h_{11}= h_{22}, h_{12}=- h_{21}$,  and the same relations hold for matrix $v$ determined from the integral equation (\ref{inteq}) and related to $\psi $ by (\ref{psiv}).

Let us introduce functions
\begin{equation} \nonumber
\left(
                                                                 \begin{array}{cc}
                                                                  a  & b \\
                                                                   -b& a \\
                                                                 \end{array}
                                                               \right)
= h_{0}(k+\alpha,k),
\end{equation}
and note that
\begin{eqnarray} \nonumber
{b}(\alpha,k)  =  \frac{1}{(2\pi)^2} \int_{\mathbb C} e^{-\overline{\alpha }z/2} e^{-i(k \overline{z} +\overline{k}z)/2} q_0(z){v}_{11}(z,k)d\sigma_z, \\  \nonumber
{a}(\alpha,k)  =  \frac{1}{(2\pi)^2} \int_{\mathbb C}   e^{-\overline{\alpha }z/2} q_0(z)\overline{v}_{12}(z,k)d\sigma_z.
\end{eqnarray}

Now define
\begin{equation} \nonumber
\left(
                                                                 \begin{array}{cc}
                                                                  {a}(\alpha,k,t)  & {b}(\alpha,k,t) \\
                                                                   -{b}(\alpha,k,t)& {a}(\alpha,k,t) \\
                                                                 \end{array}
                                                               \right)
= h(k+\alpha,k,t),
\end{equation}
where $h$ is given by (\ref{2506B}).
Similar quantities $\widehat{a},\widehat{b}$ are defined via the solutions $v(\cdot,k,t)$:
\begin{eqnarray} \nonumber
\widehat{b}(\alpha,k,t)  :=  \frac{1}{(2\pi)^2} \int_{\mathbb C}  e^{-\overline{\alpha }z/2} e^{-i(k \overline{z} +\overline{k}z)/2} q(z,t){v}_{11}(z,k,t)d\sigma_z, \\   \nonumber
\widehat{a}(\alpha,k,t)  :=  \frac{1}{(2\pi)^2} \int_{\mathbb C}    e^{-\overline{\alpha }z/2}q (z,t)\overline{v}_{12}(z,k,t)d\sigma_z.
\end{eqnarray}
These quantities are well defined due to Theorem \ref{ttttl}. Let
$$
\widehat{h}= \left(
                                                                 \begin{array}{cc}
                                                                  \widehat{a}(\alpha,k,t)  & \widehat{b}(\alpha,k,t) \\
                                                                   -\widehat{b}(\alpha,k,t)& \widehat{a}(\alpha,k,t) \\
                                                                 \end{array}
                                                               \right).
$$

Consider solution $\psi(z,k,t)$ of (\ref{fir}) with potential $Q_0$ replaced by $Q_t$, and let $v$ be defined by (\ref{psiv}). From (\ref{psiv}) it follows that $v \rightarrow I$ uniformly on each compact with respect to the variable $z$ when $k \rightarrow \infty$. Therefore, from Theorem \ref{ttttl} it follows that
\begin{equation}\label{173003A}
\widehat{a}(\alpha,k,t) \rightarrow 0, ~ k \rightarrow \infty.
\end{equation}
Obviously (see (\ref{2506B})), the same relation holds for $a(\alpha,k,t)$.

The $\overline{\partial}$-equation (\ref{dbar}) implies that the following rules are valid when $t=0$:
\begin{equation}\label{2106M2}
\frac{\partial b}{\partial \overline{k}} = \frac{\partial b}{\partial \overline{\alpha}}+ab_0, \quad \frac{\partial a}{\partial {k}} = b\overline{b_0}, \quad {\rm where} \quad b_0={b}(0,k),  ~~|\alpha|\leq A, ~~k\in \mathbb C.
\end{equation}
Due to Theorem \ref{ttttl}, the same relations are valid for $\widehat{a}(\alpha,k,t),\widehat{b}(\alpha,k,t)$:
\begin{equation}\label{2106M2a}
\frac{\partial \widehat{b}}{\partial \overline{k}} = \frac{\partial\widehat{ b}}{\partial \overline{\alpha}}+\widehat{a}\widehat{b}_0, \quad \frac{\partial \widehat{a}}{\partial {k}} = \widehat{b}\overline{\widehat{b}_0},~~~\widehat{b}_0=\widehat{b}(0,k,t), \quad  |\alpha|\leq A, ~~k\in \mathbb C,~~ 0\leq t\leq T.
\end{equation}
From (\ref{2506B}), (\ref{2106M2}), and the obvious relations
\begin{equation*}
e^{-t(\overline{k}^2-(\overline{k+\alpha})^2)/2}=e^{-t(k^2-(\overline{k+\alpha})^2)/2}\overline{e^{-i\Im k^2}}, \quad  \frac{\partial}{\partial \overline{\alpha}}e^{-t(k^2-(\overline{k+\alpha})^2)/2}=\frac{\partial}{\partial \overline{k}}e^{-t(k^2-(\overline{k+\alpha})^2)/2},
\end{equation*}
it follows that (\ref{2106M2a}) holds for $a(\alpha,k,t),b(\alpha,k,t)$:
\begin{equation}\label{2106M2b}
\frac{\partial b}{\partial \overline{k}} = \frac{\partial b}{\partial \overline{\alpha}}+ab_0, \quad \frac{\partial a}{\partial {k}} = b\overline{b_0}, \quad b_0={b}(0,k,t),  ~~|\alpha|\leq A, ~~k\in \mathbb C,~~0\leq t\leq T.
\end{equation}

Now we note that $\widehat{b}(0,k,t)=b(0,k,t)$ (see \cite[Theorem 5.3]{sung3}).  The second relations in (\ref{2106M2a}), (\ref{2106M2b})  with $\alpha=0$ imply that $(\widehat{a}-a)|_{\alpha=0}$ is anti-analytic in $k$. Then the maximum principle, together with (\ref{173003A}) for both $\widehat{a}$ and $a$, imply that $\widehat{a}|_{\alpha=0}=a|_{\alpha=0}$. Now from the first relations in (\ref{2106M2a}), (\ref{2106M2b}),  with $\alpha=0$, it follows that $\frac{\partial \widehat{b}}{\partial \overline{\alpha}}|_{\alpha=0}=\frac{\partial b}{\partial \overline{\alpha}}|_{\alpha=0}$. Then we differentiate the second relations in (\ref{2106M2a}), (\ref{2106M2b}) in $\overline{\alpha}$ and put $\alpha=0$ there. This leads to the anti-analyticity in $k$ of $\frac{\partial \widehat{a}}{\partial \overline{\alpha}}|_{\alpha=0}-\frac{\partial a}{\partial \overline{\alpha}}|_{\alpha=0}$. The maximum principle with (\ref{173003A}) imply that $\frac{\partial \widehat{a}}{\partial \overline{\alpha}}|_{\alpha=0}=\frac{\partial a}{\partial \overline{\alpha}}|_{\alpha=0}$. After the differentiation in $\overline{\alpha}$ of the first relations in (\ref{2106M2a}), (\ref{2106M2b}), we obtain that $\frac{\partial^2 \widehat{b}}{\partial \overline{\alpha}^2}|_{\alpha=0}=\frac{\partial b^2}{\partial \overline{\alpha}^2}|_{\alpha=0}$, and so on. Hence all the derivatives in $\overline{\alpha}$ of the vectors $(\widehat{a},\widehat{b})$ and $(a,b)$ coincide when $\alpha=0.$  Since both vectors are anti-analytic in $\alpha$, they are identical, i.e., $\widehat{h}=h, ~t>0$.

The {\it existence of $\widehat{v}^+$} can be shown similarly to the proof of same statement in \cite{lbcond}, where the potential was assumed to be compactly supported.
Namely, consider the following analogue of the Lippmann-Schwinger equation with different values $k_0, k\in \overline{D}$ of the spectral parameter in the operator and in the free term of the equation:
\begin{eqnarray}\label{173003B}
\psi^+(z,k)=e^{i\frac{\overline{k}z}{2}}I+ \int_{z \in \mathbb C} G(z-z',k_0) Q_t(z') \overline{\psi^+}(z',k) d\sigma_{z'}, \quad G(z,k)=\frac{1}{\pi}\frac{e^{i\overline{k}z/2 }}{z}.
\end{eqnarray}
We substitute here $\psi^+=\mu^+ e^{i\overline{k_0}z/2}$ and rewrite the equation in terms of
\begin{equation}\label{172604A}
w^+=\mu^+(z, k)-e^{i(\overline{k-k_0})z/2}I \in L^\infty_z(L^p_k), ~p>1.
\end{equation}
The equation takes the form
\begin{eqnarray}\label{173003Bmu}
w^+(z,k)-\int_{z \in \mathbb C} \frac{e^{-i\Re(\overline{k_0}z')}}{z-z'} Q_t(z') \overline{w^+}(z',k) d\sigma_{z'} =
\int_{z \in \mathbb C} \frac{e^{-i\Re(\overline{k_0}z')}}{z-z'} \left [Q_t(z') e^{-i\overline{z'}(k-k_0)/2}\right ]d\sigma_{z'}.
\end{eqnarray}

   Theorem \ref{ttttl} implies that function $\left [Q_t(z') e^{-i\overline{z'}(k-k_0)/2}\right ]$ decays exponentially as $z\to\infty$, and $|k|, |k_0|\leq A$. The unique solvability of the problem (\ref{173003Bmu}) is obvious since the potential is small.

Function $\widehat{v}^+$ is defined by $\psi^+$ in the same way as $v$ is defined by $\psi$ in (\ref{psiv}). The analyticity of $\widehat{v}^+$ and (\ref{170204B}) are proved in Lemmas 3.1 and  3.5 of \cite{lbcond}.

\qed

\section{Proof of statement 2 of the Theorem \ref{t111}.}

 {\bf Reduction to Theorem \ref{ttll} and Lemma \ref{lll}.} Theorem \ref{ttttl} immediately implies that the operator $T_{z,t}:\mathcal B^s\to\mathcal B^s, ~s>2,$ is analytic in $x$ and $y$ in a complex neighborhood of $\mathbb R^2$. In order to use the {multidimensional} analytic Fredholm theory (\cite[Th. 4.11, 4.12]{mt2} or \cite{mt}) and obtain a decay of operator norm $\|T^2_{z,t}\|$ as $|z|\to\infty$, we would like to consider this operator in the Hilbert space $\mathcal B^2$ instead of the Banach space $\mathcal B^s, ~s>2$. All the previous and new results mentioned in this paper remain valid if $s>2$ is replaced by $s=2$ (with the appropriate definition of the space $\mathcal B^2$ given in (\ref{170904AZ})). In order to justify the latter statement, one needs to show that the properties of the operator $T_{z,t}$ are preserved when $s>2$ is replaced by $s=2$. This will be done in Theorem \ref{ttll} below (we will not discuss the  properties that obviously are $s$-independent), but we will show that operator $T_{z,t}:\mathcal B^2\to \mathcal B^2, 0\leq t\leq T,$ is compact, continuous in $(z,t)$, and analytic in $(x,y)$ in a complex neighborhood of  $\mathbb R^2$.
After that, we will show (Lemma \ref{lll}) the invertibility of $I+T_{z,t}$ at large values of $|z|$. Then the second statement of the theorem will be a simple consequence of the first statement and the analytic Fredholm theory. Note that the invertibility of operator $I+T_{z,t}$ will be proved for $z$ on each ray $\arg z=\psi =const, |z|\geq Z_0,$ with $\psi$-independent $Z_0$ and with $T_{z,t}$ defined (see (\ref{0712G})) using a special value of $k_0=k_0(\psi)$. Since the solution  $(q,\phi)$ of problem (\ref{dsii}) does not depend on the choice of $k_0$ (see Theorem \ref{t111}), it remains only to prove Theorem \ref{ttll} and Lemma \ref{lll}.
\subsection{Compactness of operator $T$}
We will need the following lemma.
\begin{lemma}\label{coc}
Let operator $M:\mathcal B^2\to \mathcal B^2$ have the form
\[
(Mf)(k)=\int_{\mathbb C\backslash D}   \frac{g(\varsigma)}{\varsigma-k}f(\varsigma)d\sigma_\varsigma, \quad k\in \mathbb C,
\]
where function $g_\delta=g(\varsigma)(1+|\varsigma|)^\delta$has the following properties
\[
|g_\delta|<a_1<\infty,~ g_\delta\to 0~~  as ~~\varsigma\to\infty,~~ and ~~\|g_\delta\|_{L^2(\mathbb C\backslash D)}=a_2<\infty
 \]
  for some $\delta>0$. Then $M$ is compact and $\|M\|\leq C(a_1+a_2)$.
 \end{lemma}
{\bf Proof.} Let $P$ be the following operator in $\mathcal B^2$ of rank one:
\begin{equation}\label{290118A}
Pf=-\frac{\beta(k)}{k}\int_{\mathbb C\backslash D}g(\varsigma)f(\varsigma)d\sigma_\varsigma,
\end{equation}
where $\beta$ is the function introduced in the definition of the space $\mathcal B^2$. Since $Pf=0$ in a neighborhood of $D$, and
\[
\int_{\mathbb C\backslash D}g(\varsigma)f(\varsigma)d\sigma_\varsigma\leq a_2\int_{\mathbb C\backslash D}|\frac{f(\varsigma)}{(1+|\varsigma|)^\delta}|^{2}d\sigma_\varsigma\leq Ca_2\|f\|_{\mathcal B^2},
\]
it is enough to prove the statement of the lemma for operator
$M-P=M_1+M_2$, where
\[
M_if=\int_{\mathbb C\backslash D}K_i(k,\varsigma)f(\varsigma)d\sigma_\varsigma, ~~ K_1(k,\varsigma)=
\frac{\alpha(\varsigma -k)}{\varsigma -k}g(\varsigma), ~~ K_2(k,\varsigma)=[\frac{\beta(\varsigma -k)}{\varsigma -k}+\frac{\beta(k)}{k}]g(\varsigma),
\]
and $\alpha:=1-\beta$ is a cut-off function which is equal to one in a neighborhood of $D$.

Let $M_i'$ be the operator defined by the same formulas as operators $M_i$, but considered as operators in $L^2(\mathbb C)$. Let us show that operators $M_i'$ are compact and their norms do not exceed $C(a_1+a_2)$.

Since $|g|\leq a_1$, we have
\[
\sup_{k\in \mathbb C}\int_\mathbb C |K_1(k,\varsigma)|d\sigma_\varsigma+\sup_{\varsigma\in \mathbb C}\int_\mathbb C |K_1(k,\varsigma)|d\sigma_k\leq C a_1.
\]
Hence, from the Young theorem, it follows that $\|M_i'\|\leq Ca_1$.
Similarly, using the decay of $g_1$ at infinity, we obtain that $M_1'=\lim_{R\to\infty} M_{1,R}'$, where $M_{1,R}'$ are operators in $L^2(\mathbb C)$ with the integral kernels $K_1(k,\varsigma)\alpha(\varsigma/R)$. Operators $M_{1,R}$ are pseudo-differential operators of order $-1$ (they increase the smoothness of functions by one) defined in a bounded domain.
Hence operators $M_{1,R}'$ and their limit $M_{1}'$ are compact operators in $L^2(\mathbb C)$.

The boundedness (with the upper bound $Ca_2$) and compactness of the operator $M_2$ will be proved if we show that
\[
\int_\mathbb C\int_\mathbb C |K_2(k,\varsigma)| ^2d\sigma_kd\sigma_\varsigma\leq Ca_2.
\]
We split the interior integral in two parts: over region $|k|<2|\varsigma|$ and over region $|k|>2|\varsigma|$, and estimate each of them separately. We have
\[
\int_{|k|<2|\varsigma|} |K_2(k,\varsigma)| ^2d\sigma_k\leq 2|g(\varsigma)|^2\int_{|k|<2|\varsigma|} [\frac{\beta^2(\varsigma -k)}{|\varsigma -k|^2}+\frac{\beta^2(k)}{|k|^2}] d\sigma_k\leq C|g(\varsigma)|^2(1+|\varsigma|)^{\delta}.
\]
A better estimate with a logarithmic factor is valid, but we do not need this accuracy. Next,
\[
\int_{|k|>2|\varsigma|} |K_2(k,\varsigma)| ^2d\sigma_k= |g(\varsigma)|^2\int_{|k|>2|\varsigma|}
\frac{|k\beta(\varsigma -k)+(\varsigma -k)\beta(k)|^2}{|(\varsigma -k)k|^2}d\sigma_k.
\]
The denominator of the integrand can be estimated from below by $\frac{1}{4}|k|^4$. The numerator, denoted by $n$, has the following properties. If $|\varsigma|$ is large enough, than both beta functions in $n$ are equal to one, and $n=|\varsigma|^2$. The same is true if $|\varsigma|$ is bounded and $|k|$ is large. If both variables are bounded, than $|n|$ is bounded. Thus $|n|<(C+|\varsigma|)^2$, and the integrand above does not exceed $C\frac{1+|\varsigma|^2}{|k|^4}$. Obviously, the integrand vanishes when $|k|$ is small enough. Thus there is a constant $c>0$ such that
\[
\int_{|k|>2|\varsigma|} |K_2(k,\varsigma)| ^2d\sigma_k\leq C|g(\varsigma)|^2\int_{|k|>\max(c,2|\varsigma|)}\frac{1+|\varsigma|^2}{|k|^4}d\sigma_k
\]
\[
\leq C|g(\varsigma)|^2\int_{|k|>c}\frac{1}{|k|^4}d\sigma_k+
C|g(\varsigma)|^2\int_{|k|>2|\varsigma|}\frac{|\varsigma|^2}{|k|^4}d\sigma_k=C_1|g(\varsigma)|^2.
\]
Hence
\[
\int_\mathbb C\int_\mathbb C |K_2(k,\varsigma)| ^2d\sigma_kd\sigma_\varsigma\leq C
\int_{\mathbb C} |g(\varsigma)| ^2(1+|\varsigma|)^\delta d\sigma_\varsigma\leq C'a_2.
\]
Thus, operators
$M_i':L^2(\mathbb C)\to L^2(\mathbb C)$ are compact and $\|M_i'\|\leq C(a_1+a_2)$.

Denote by  $M_i'':\mathcal B^2\to L^2(\mathbb C)$ operators with the same integral kernels $K_i$ as for operators $M_i'$, but with the domain $\mathcal B^2$ instead of $L^2(\mathbb C)$. Compactness of these operators will be proved if we show the boundedness of $M_i'$ on the one-dimensional space of functions of the form $f_c(\varsigma)=c\frac{\beta(\varsigma)}{\varsigma},~c=$const. The upper estimate on $\|M_i''f_c \|$ can be obtained by repeating the arguments above used to estimate $\|M_i'\|$. One needs only to replace $f_c$ by the function $f=f_c/|\varsigma|^{\delta/2}\in L^2(\mathbb C)$ and replace the kernel $K_i$ by $K_i|\varsigma|^{\delta/2}$. Hence, operators $M_i''$ are compact and $\|M_i'\|\leq C(a_1+a_2)$.

Obviously, for each $f\in \mathcal B^2$, the function $(M_1+M_2)f$ is analytic in $D$. Consider its trace on $\partial D$. Let $M_{D}:\mathcal B^2\to L^2(\partial D)$ be the operator that maps each $f\in \mathcal B^2$ into the trace of $(M_1+M_2)f$ on $\partial D$. In order to complete the proof of the lemma, it remains to show that operator $M_{D}$ is well defined, compact, and $\|M_{D}\|\leq C(a_1+a_2)$. To prove these properties of $M_{D}$, we split the operator into two terms $M_{D}=M_{D}\phi+M_{D}(1-\phi)$, where $\phi$ is the operator of multiplication by the indicator function of a disk $D_1$ of a larger radius than the radius of $D$. Then $M(1-\phi)f$ is analytic in $D_1$, and
\[
\|M(1-\phi)f\|_{L^2(D_1)}\leq \|Mf\|_{L^2(\mathbb C)}\leq C(a_1+a_2)\|f\|_{\mathcal B^2}.
\]

From a priori estimates for elliptic operators, it follows that
\[
\|M(1-\phi)f\|_{H^s(D)}\leq C_s\|M(1-\phi)f\|_{L^2(D_1)}\leq C_s(a_1+a_2)\|f\|_{\mathcal B^2},
 \]
 where $H^s$ is the Sobolev space and $s$ is arbitrary. Hence $\|M(1-\phi)f\|_{H^{s-1/2}(\partial D)}\leq C(a_1+a_2)\|f\|_{\mathcal B^2}$. This implies that operator $M_D(1-\phi)$ is compact and  its norm does not exceed $C(a_1+a_2)$. We will take $D_1$ not very large, so that function $\beta$ vanishes on $D_1$. Then $M\phi f$ is the convolution of $1/k$ and $\phi gf$, i.e., $M\phi f=\frac{1}{k}*(\phi gf)$. The latter expression  is a pseudo differential operator of order $-1$ applied to the function $\phi gf$ with a compact support. Thus,
 \[
 \|M\phi f\|_{H^1(D)}\leq C\|\phi gf\|_{L^2(D_1)}\leq Ca_1\|f\|_{\mathcal B^2},
 \]
and therefore $\|M_D\phi f\|_{H^{1/2}(D)}\leq Ca_1\|f\|_{\mathcal B^2}$. Hence, operator $M_D\phi$ is compact and its norm does not exceed $Ca_1$.

 \qed

\begin{theorem} \label{ttll} Let conditions of Theorem \ref{t111} hold. Then
operator $T_{z,t}:\mathcal B^2\to \mathcal B^2, 0\leq t\leq T,$ is compact, continuous in $(z,t)$, and analytic in $(x,y)$ in a complex neighborhood of  $\mathbb R^2$. The same properties are valid for derivatives of $T_{z,t}$  of any order in $t,x,y$.
\end{theorem}
{\bf Remark.} $T_{z,t}$ is analytic in $x,y$ in the region $|\Im x|^2+|\Im y|^2\leq R^2$.

{\bf Proof.}  The operator $T_{z,t}$ can be naturally split into two terms: $T_{z,t}=\m + \D$, where $\m$ involves integration over $\mathbb C \backslash D$ and $\D$ involves integration over $\partial D$. In particular,
$$
 \m\phi = \frac{1}{\pi} \int_{\mathbb C \backslash D} \frac{e^{i \Re (\varsigma \overline{z}) } \overline{\phi}(\varsigma)\Pi^oh(\varsigma,\varsigma,t)}{ \varsigma - k } d\sigma_\varsigma.
$$
The statements of the theorem are valid for operator $\m$ due to (\ref{sm}), Lemma \ref{coc} and Theorem~\ref{tttt}. Indeed, the compactness and continuity of $M$ in $(z,t)$ is proved in Lemma \ref{coc}. The analyticity  in $(x,y)$ follows from the fast decay of $h$ at infinity which is established in Theorem~\ref{tttt}.

Let us show that the same properties are valid for $\D$. We write $\D$ in the form $D=I_1I_2$, where operator $I_2 ~:~ L^2(\partial D) \rightarrow C^{\alpha}(\partial D)$ is defined by the interior integral in the expression for $\D$ in (\ref{0712G}), and operator $I_1:C^{\alpha}(\partial D)\to \mathcal H^s$ is defined by the exterior integral in the same expression.  Here $C^{\alpha}(\partial D)$ is the Holder space and  $\alpha$ is an arbitrary number in $(0,1/2)$.  The integral kernel of operator $I_2$ has a logarithmic singularity at $\varsigma=\varsigma'$, i.e., $I_2$ is a pseudo differential operator of order $-1$, and therefore $I_2$ is a bounded operator from $L^2(\partial D)$ into the  Sobolev space $H^1(\partial D)$. Thus it is compact as operator from $C(\partial D)$ to $C^\alpha(\partial D), \alpha \in (0,1/2)$, due to the Sobolev embedding theorem. Thus the compactness of  $\D$ will be proved as soon as we show that $I_1$ is bounded.

For each $\phi \in C^\alpha(\partial D)$, function $I_1\phi$ is analytic outside of $\partial D$ and vanishes at infinity. Due to the Sokhotski–Plemelj theorem, the limiting values $(I_1\phi)_\pm$ of $(I_1\phi)$ on $\partial D$ from inside and outside of $D$, respectively, are equal to $\frac{\pm\phi}{2}+P.V.\frac{1}{2\pi i} \int_{\partial D} \frac{\phi(\varsigma)d\varsigma}{\varsigma - \lambda}$. Thus
\[
\max_{\partial D}|(I_1\phi)_\pm|\leq C\|\phi\|_{C^\alpha(\partial D)}.
\]
From the maximum principle for analytic functions, it follows that the same estimate is valid for function $I_1\phi$ on the whole plane. Taking also into account that $I_2\phi$ has the following behavior at infinity $I_2\phi\sim c/k+O(|k|^2)$, we obtain that  operator $I_1$ is bounded. Hence operator $\D$ is compact. Since  $h$ decays superexponentially at infinity, the arguments above allow one to obtain not only the compactness of $\D$, but also its smoothness in $t,x,y$ and analyticity in $(x,y)$.

\qed

\subsection{The invertibility of $I+T_{z,t}$ at large values of $z$}
We will prove the following lemma.
\begin{lemma} \label{lll}
The following relation is valid for operator norm of $T_{z,t}^2$ in $\mathcal B^2$:
\begin{equation} \nonumber
\max_{0\leq t\leq T}\|T_{z,t}^2\|\to 0,   \quad z\in \mathbb C,~ z\to\infty.
\end{equation}
Hence the operator $I+T_{z,t}$ is invertible when $z\in \mathbb C,~|z|\gg 1$.
\end{lemma}
We split operator $T_{z,t}$ into two terms $T_{z,t}=\m+\mathcal D$ that correspond to the integration over $\mathcal C\backslash D$ and $D$, respectively, in (\ref{0712G}).
The entries $M^{ij}, ~D^{ij}, ~i,j=1,2,$ of the matrix operators $\m$ and $\D$ are
$$
M^{ 11}=M^{22}=0,  \quad  M^{12} \phi =-M^{21} \phi = \frac{1}{\pi} \int_{\mathbb C \backslash D} \frac{e^{i \Re (\varsigma \overline{z} ) -t(\varsigma^2-\overline{\varsigma}^2)/2} \overline{\phi}(\varsigma)h_{12}(\varsigma,\varsigma)}{ \varsigma - k } d\sigma_\varsigma,
$$
$$
D^{11} \phi=D^{22} \phi    = \frac{1}{2\pi i}\int_{\partial D}\frac{d\zeta}{\zeta -k}\int_{\partial D}
\overline{{\rm Ln}\frac{\overline{\varsigma'}-\overline{\varsigma} }{\overline{\varsigma'}-\overline{k_0} } h_{11}(\varsigma',\varsigma) }e^{\frac{i}{2}(\varsigma-\varsigma')\overline z+\frac{t}{2}(\varsigma'^2-\varsigma^2)} \phi(\varsigma '){d\zeta'},
$$
\begin{equation} \nonumber
D^{12}\phi = -D^{21}\phi=\frac{1}{2\pi i} \int_{\partial D} \frac{d\zeta}{\zeta -k}\int_{\partial D} {\rm Ln}\frac{\overline{\varsigma'}-\overline{\varsigma} }{\overline{\varsigma'}-\overline{k_0} }
  {h_{12}(\varsigma',\varsigma)} e^{{\frac{i}{2}}(\varsigma \overline{z}+\overline{\varsigma'}z)+\frac{t}{2}(\overline{\varsigma'}^2-\varsigma^2)}\overline{\phi}(\varsigma ')\overline{d\varsigma'}.
\end{equation}
We used here the relations $h_{12}=-h_{12}, h_{11}=h_{22}$ for the entries of $h_0$ that were established, for example, in \cite[Lemma 4.1]{lvds}.

Lemma \ref{coc} implies the uniform boundedness of $M^{21},M^{12}$ when $0\leq t \leq T,~z \in \mathbb C$. Thus Lemma \ref{lll} will be proved if we show that operator norms of
$M^{21}\overline{M^{12}}$ and  $D^{ij}, i,j=1,2$, vanish as $z \to \infty$. Let us prove the statement about $D^{ij}$.

\begin{lemma}\label{171704Ba}
 For each $T>0$, there exists a constant $C_{T}$ such that
$$
\|D \varphi\|_{\mathcal B^2} \leq \frac{C_{\alpha,T}}{1+|z|^{1/4}} \|\varphi \|_{\mathcal B^2}, \quad z \in \mathbb C,~~0\leq t\leq T,
$$
if $k_0$ in the definition of operator $\mathcal D$ is chosen to belong to $\partial D$ and equal to $k_0 =-iAe^{i\psi}$, where $\psi=\arg z$ and $A$ is the radius of the disk $D$.
\end{lemma}

{\bf Proof.} We will prove the estimate for the component $D^{12}$ of the matrix $D$. Other components of $D$ can be estimated similarly. Consider the interior integral in $D^{12}$:
\begin{equation}\label{171704Abc}
R^{12}\phi=\int_{\partial D} {\rm Ln}\frac{\overline{\varsigma'}-\overline{\varsigma} }{\overline{\varsigma'}-\overline{k_0} }
  {h_{12}(\varsigma',\varsigma)} e^{{\frac{i}{2}}(\varsigma \overline{z}+\overline{\varsigma'}z)+\frac{t}{2}(\overline{\varsigma'}^2-\varsigma^2)}\overline{\phi}(\varsigma ')d\overline{\varsigma'} ,\quad  \varsigma \in \partial D,~~\phi\in \mathcal B^2.
\end{equation}
Our goal is to show that
\begin{equation}\label{fg}
\|R^{12}\phi\|_{L^\infty(\partial D)}\leq \frac{C_{T}}{1+|z|^{1/4}} \|\phi \|_{L^2(\partial D)} ,\quad~\phi\in \mathcal B^2.
\end{equation}

 The integrand in (\ref{171704Abc}) is anti-holomorphic in $\varsigma'\in D$ with logarithmic branching points at $k_0$ and $\varsigma$.
 If $k_0$ is strictly inside $D$, then the integration  over $\partial D$ in (\ref{171704Abc}) can be replaced by the integration over two sides of the segment $[k_0,\varsigma]$, which are passed in the counter clock-wise direction. The values of the logarithm on these sides differ by the constant $2\pi$. This leads to an alternative form of the operator $\mathcal D$:
\begin{equation}\nonumber
D^{12}\phi = -D^{21}\phi=i\int_{\partial D} \frac{d\zeta}{\zeta -k}\int_{[k_0,\varsigma]}
  {h_{12}(\varsigma',\varsigma)} e^{{\frac{i}{2}}(\varsigma \overline{z}+\overline{\varsigma'}z)+\frac{t}{2}(\overline{\varsigma'}^2-\varsigma^2)}\overline{\phi}(\varsigma ')\overline{d\varsigma'}.
\end{equation}
If $k_0\in \partial D$, the contour of integration above can be replaced by  arc$[k_0,\varsigma]$. Thus
\begin{equation} \nonumber
R^{12}\phi=i\int_{\widehat{k_0,\varsigma}}
  {h_{12}(\varsigma',\varsigma)} e^{{\frac{i}{2}}(\varsigma \overline{z}+\overline{\varsigma'}z)+\frac{t}{2}(\overline{\varsigma'}^2-\varsigma^2)}\phi(\varsigma ')d\overline{\zeta'} ,\quad  \varsigma \in \partial D,~~\phi\in L^2(\partial D).
\end{equation}

Consider the following function (from the exponent in the integrand above): $\Phi=\Re \left[\frac{i}{2}\varsigma \overline{z}\right].$  This function is linear in $\varsigma$, and for each fixed $z=|z|e^{i\psi}, \psi \in[0,2\pi)$, it has the unique global maximum on $D$. The maximum occurs on the boundary at the point $\varsigma_0 =-iAe^{i\psi}$, which depends only on the argument of $z$. Due to Theorem \ref{t111}, point $k_0\in \partial D$ can be chosen arbitrarily.
 We choose $k_0 =\varsigma_0 \in \partial D$, and we get that
$$
|R^{12}\phi| \leq C\left (\int_{\widehat{\varsigma_0,\varsigma}}  \exp 2 \left(\Phi(\varsigma)-\Phi(\varsigma')\right )|d\varsigma '| \right)^{1/2}\|\phi \|_{L^2}. \quad
$$
Let us estimate the integral above. Let $\varsigma=-iAe^{i(\psi+\varphi)},~|\varphi|\leq \pi$. For $\varsigma'\in  \widehat{\varsigma_0, \varsigma}$, we have
$$
\Phi(\varsigma')=A|z|(\cos\varphi')/2, \quad \Phi(\varsigma)=A|z|(\cos\varphi)/2,
$$
and the integral is equal to
$$
\int_0^\varphi e^{A|z|(\cos \varphi -\cos \varphi')/2}d\varphi'= O(\frac{1}{\sqrt{|z|}}), ~ z \rightarrow \infty.
$$
This justifies (\ref{fg}).

\qed

Let us show now that the following statement holds.
\begin{lemma}\label{310118A}
\begin{equation}\label{oin2}
\max_{0\leq t\leq T}\|M^{21}\overline{M^{12}}\|_{\mathcal B^2}\to 0,   \quad z\in \mathbb C,~ z\to\infty.
\end{equation}
\end{lemma}
{\bf Proof.} Kernels of $M^{12},M^{21}$ are smooth, see (\ref{sm} ). From Theorem \ref{tttt}, it follows that the kernels and rapidly decaying functions in $\mathbb C$. Therefore, Lemma \ref{coc}
implies that operators $M^{12},M^{21}$ can be approximated in $\mathcal B^2$ by operators with function $h_{12}$ replaced by a compactly supported one. Therefore, without loss of the generality, we will assume below that the supports of $h_{12},h_{21}$ belong to a bounded domain $\mathcal O$.

We will use the notation $P$ for the one-dimensional operator defined in (\ref{290118A}) with the density $g={e^{i \Re (\varsigma \overline{z}) -t(\varsigma-\overline{\varsigma}^2)/2} h_{12}(\varsigma,\varsigma)}$. Let $\widehat{M}:=(M^{12}-P)\overline{(M^{21}-P)}$.  We will prove that
\begin{equation}\label{oin2a}
\max_{0\leq t\leq T}\|\widehat{M}\|_{\mathcal B^2}\to 0,   \quad z\in \mathbb C,~ z\to\infty.
\end{equation}
The other three terms $M^{12}(\overline{M^{21}-P)},~(M^{12}-P)\overline{M^{21}}$, and $P\overline{P}$ can be treated in the same way. We have
$$
\widehat{M}\varphi = \frac{1}{\pi^2} \int_{\mathcal O\backslash D} A(z,\varsigma,\varsigma_2) \overline{h_{21}}(\varsigma_2,\varsigma_2) e^{-i \Re (\varsigma_2 \overline{z}) +t(\varsigma_2-\overline{\varsigma_2}^2)/2}\varphi(\varsigma_2) d\sigma_{\varsigma_2},
$$
where
\begin{equation}\label{AAAa}
A(z,\varsigma,\varsigma_2) := \int_{\mathcal O \backslash D}{e^{i \Re (\varsigma_1 \overline{z}) -t(\varsigma_1-\overline{\varsigma_1}^2)/2} h_{12}(\varsigma_1,\varsigma_1)}
\left (\frac{1}{\varsigma_1 - \varsigma} +\frac{\beta(\varsigma)}{\varsigma} \right ) \overline{\left (\frac{1}{\varsigma_2 - \varsigma_1} +\frac{\beta(\varsigma_1)}{\varsigma_1} \right )} d\sigma_{\varsigma_1}.
\end{equation}
The Minkovsky inequality in the integral form implies the following two estimates.
\begin{eqnarray}\nonumber
\|\widehat{M}f\|_{L^2(\mathbb C \backslash D)}  \leq  \int_{\mathcal O \backslash D}\left [ \int_{\mathcal O \backslash D} |A(z,\varsigma,\varsigma_2)|^2   d\sigma_{\varsigma} \right ]^{1/2} |h_{21}(\varsigma_2,\varsigma_2) f(\varsigma_2)|d\sigma_{\varsigma_2}, ~~f \in \mathcal B^2.
\\ \nonumber
\|\widehat{M}f\|_{L^2(\partial  D)}  \leq  \int_{\mathcal O \backslash D}\left [ \int_{\partial D} |A(z,\varsigma,\varsigma_2)|^2   |d{\varsigma}| \right ]^{1/2} |h_{21}(\varsigma_2,\varsigma_2) f(\varsigma_2)|d\sigma_{\varsigma_2}, ~~f \in \mathcal B^2.
\end{eqnarray}
Since the norm of the operator $L^2(\mathbb C \backslash D) \to L^1(\mathbb C \backslash D)$ of multiplication by $h_{21}$ can be estimated by a constant, the validity of (\ref{oin2a}) will follow from the estimates above if we show that
\begin{equation} \nonumber
\sup_{\varsigma_2 \in \mathbb C \backslash D}\int_{\mathcal O \backslash D} |A(z,\varsigma,\varsigma_2)|^2 d\sigma_{\varsigma} \rightarrow 0, \quad \sup_{\varsigma_2 \in \mathbb C \backslash D}\int_{\partial D} |A(z,\varsigma,\varsigma_2)|^2 |d{\varsigma}| \rightarrow 0, \mbox{ as } z \rightarrow \infty.
\end{equation}

We will prove only the first inequality above, since the second one can be proved similarly.
Note that,
uniformly in $\varsigma_2 \in \mathcal O$,
\[
\int_{\mathcal O \backslash D} |A(z,\varsigma,\varsigma_2)|^2 d\sigma_{\varsigma}
\]
\begin{equation}\nonumber
\leq \int_{\mathcal O \backslash D}\left| \int_{\mathcal O\backslash D} {h_{12}(\varsigma_1,\varsigma_1)}\left (\frac{1}{\varsigma_1 - \varsigma} +\frac{\beta(\varsigma)}{\varsigma} \right ) \overline{\left (\frac{1}{\varsigma_2 - \varsigma_1} +\frac{\beta(\varsigma_1)}{\varsigma_1} \right )}  d{\sigma_{\varsigma_1}} \right |^2 d{\sigma_{\varsigma}} < C.
\end{equation}
The boundedness follows from the fact that the internal integral is $O(\ln|\varsigma-\varsigma_2|), \varsigma-\varsigma_2 \rightarrow 0.$ Let $A^{s}$ be given by (\ref{AAAa}) with the extra factor $\eta_s:=\eta(s|\varsigma-\varsigma_1|)\eta(s|\varsigma_1-\varsigma_2|)), s>0$, in the integrand, where $\eta\in C^\infty(\mathbb R),~ \eta=1$ outside of a neighborhood of the origin, and $\eta$ vanishes in a smaller neighborhood of the origin.


For each $\varepsilon$, there exists $s=s_0(\varepsilon)$ such that
\[
\int_{\mathcal O \backslash D}|A-A^{s_0}|^2\, d{\sigma_{\varsigma}}<\varepsilon
\]
for all the values of $\varsigma_2\in\mathcal O,z\in\mathbb C$. Denote by $R^{s_0}$ the function $A^{s_0}$ with the potential ${h}_{12}$ replaced by its $L_1$-approximation $\widetilde{h}_{12}\in C_0^\infty(\mathbb C \backslash D)$. We can choose this approximation in such a way that
\[
\int_{\mathcal O \backslash D}|A^{s_0}-R^{s_0}|^2\, d{\sigma_{\varsigma}}<\varepsilon
\]
for all the values of $\varsigma_2,z$. Now it is enough to show that
\[
|R^{s_0}(\varsigma,\varsigma_2,z)|\to 0 \quad {\rm as} \quad z\to \infty
\]
uniformly in $\varsigma,\varsigma_2 \in \mathcal O$. The latter can be obtained by integration by parts in
\begin{equation}\nonumber
R^{s_0}_z(\varsigma,\varsigma_2) := \int_{\mathbb C \backslash D}(1-\eta_s){e^{i \Re (\varsigma_1 \overline{z}) -t(\varsigma_1-\overline{\varsigma_1}^2)/2} \widetilde{h_{12}}(\varsigma_1,\varsigma_1)}
\left (\frac{1}{\varsigma_1 - \varsigma} +\frac{\beta(\varsigma)}{\varsigma} \right ) \overline{\left (\frac{1}{\varsigma_2 - \varsigma_1} +\frac{\beta(\varsigma_1)}{\varsigma_1} \right )} d\sigma_{\varsigma_1}
\end{equation}
(integrating $e^{i\Re (\varsigma_1 \overline{z})} $ and differentiating the complementary factor). This completes the proof of
(\ref{oin2}).

\qed

\end{document}